\documentclass[twocolumn,showpacs,preprintnumbers,amsmath,amssymb]{revtex4}
\usepackage{graphicx}
\usepackage{bm}
\usepackage{latexsym}
\usepackage{epsfig}
\usepackage{natbib}
\usepackage{color}

\newcommand{\CC}{\Lambda}
\newcommand{\rL}{\rho_{\CC}}
\newcommand{\rLo}{\rho_{\CC}^0}

\begin{document}

\title{Complete cosmic history with a dynamical $\Lambda=\Lambda (H)$ term}

 \author{E. L. D. Perico} \email{elduartep@usp.br}
\affiliation{Instituto de F\'{i}sica, Universidade de S\~ao Paulo, Rua do Mat\~ao travessa R, 05508-090, S\~ao Paulo, SP, Brazil}

 \author{J. A. S. Lima} \email{limajas@astro.iag.usp.br}
\affiliation{Departamento de Astronomia, Universidade de S\~ao Paulo, Rua
do Mat\~ao 1226, 05508-900, S\~ao Paulo, SP, Brazil}

\author{Spyros Basilakos} \email{svasil@academyofathens.gr}
\affiliation{Academy of Athens, Research Center for Astronomy and Applied
Mathematics, Soranou Efesiou 4, 11527, Athens, Greece}

\author{Joan Sol\`a}     \email{sola@ecm.ub.edu}
\affiliation{High Energy Physics Group, Departament d'Estructura i
Constituents de la Mat\`eria, and Institut de Ci\`encies del Cosmos (ICC),
Univ. de Barcelona, Avinguda Diagonal 647 E-08028 Barcelona, Catalonia,
Spain}

\date{\today }

\begin{abstract}
In the present mainstream cosmology, matter and spacetime emerged from a
singularity and evolved through four distinct periods: early inflation,
radiation,  dark matter and late-time inflation (driven by dark energy).
During the radiation and dark matter dominated stages, the universe is
decelerating while the early and late-time inflations are accelerating
stages. A possible connection between the accelerating periods remains
unknown, and, even more intriguing, the best dark energy candidate
powering the present accelerating stage ($\Lambda$-vacuum) is plagued with
the cosmological constant and coincidence puzzles. Here we propose an
alternative solution for such problems based on a large class of
time-dependent vacuum energy density models in the form of power series of
the Hubble rate, $\Lambda=\Lambda (H)$. The proposed class of
$\Lambda(H)$-decaying vacuum model provides: i) {a new mechanism for
inflation} ({different from} the usual inflaton models), (ii) a natural
mechanism for a graceful exit, which is universal for the {whole class of
models}; iii) the currently accelerated expansion of the universe, iv) a
mild dynamical dark energy at present; and v) a final de Sitter stage.
Remarkably, the late-time cosmic expansion history of our class of models
is very close to the concordance $\Lambda$CDM model, but above all it
furnishes the necessary smooth link between the initial and final de
Sitter stages through the radiation- and matter-dominated epochs.
\end{abstract}

\maketitle

\section{Introduction}
Several cosmological observations (supernovae type Ia, CMB, galaxy
clustering, etc.) have converged to a paradigm of a cosmic expansion
history that involves a spatially flat geometry and a recently initiated
accelerated expansion of the universe
\cite{PeeblesRatra03,KowalskiET08,HickenET09,KomatsuET09,HinshawET09,amanullah2010,KomatsuET11,LimaAlcaniz00,JesusCunha09,BasilakosPlionis10,Ade13}. This expansion
has been attributed to an energy component called dark energy (DE) with
negative pressure, which dominates the universe at late times. The easiest
way to fit the current cosmological data is to include in the Friedmann
equations the cosmological constant (CC)
\cite{BasilakosPlionis10,amanullah2010,Ade13}. Despite the fact that the
so-called concordance model (or $\Lambda$CDM model) describes well the
global properties of the observed universe it suffers from the CC problem \cite{Zee85,Weinberg89}.
However, the alternative frameworks (e.g. quintessence models and the
like) are not free from similar fine-tuning and other no less severe
problems (including the presence of extremely tiny masses). Whichever way
it is formulated, the CC problem appears as a tough issue which involves
many faces: not only the problem of understanding the tiny current value
of the vacuum energy density ($\rho_{\Lambda}=c^{2}\,\Lambda/8\pi G\simeq
10^{-47}\,GeV^4$) \cite{Weinberg89} in the context of quantum field theory
(QFT) or string theory, but also the cosmic coincidence problem, i.e. why
the density of matter is now so close to the vacuum density
\cite{Steinhardt97}.

Even before the discovery of the accelerating universe  based on
Supernovae observations (see \cite{KowalskiET08,HickenET09,KomatsuET09,
HinshawET09} and Refs. therein), a great deal of attention was dedicated
to time-evolving vacuum models, $\Lambda\equiv\Lambda(t)$, motivated
basically by the age of the universe and CC problems
\cite{OzerTaha87,FreeseET87,Bertolami86,
CarvalhoET92,LM93,Waga93,ArcuriWaga94,LM94,LT96,SalimWaga93,Arbab97} (see
also \cite{OverduinCooper98} for a short review of this earlier
literature). These models also act as an important alternative to the
cosmic concordance ($\Lambda$CDM) and scalar-fields dark energy models,
since they can explain in an efficient way the accelerated expansion of
the universe and also provide an interesting attempt to evade the
coincidence and cosmological constant problems of the standard
$\Lambda$-cosmology (see, for instance, Lima in \cite{PeeblesRatra03}).

Although the precise functional form of $\Lambda(t)$ is not known, which
is however also the case for the vast majority of the usual dark energy
models, an interesting QFT approach within the context of the
renormalization group (RG) was proposed long time ago
\cite{NelsonPanan82,ParkerToms85}. Later on, the RG-running framework was
further explored in
\cite{ShapSol00,BabicET02,ShapSol03,SolaStefancic05,Fossil07,ShapSol09}
from the viewpoint of QFT in curved spacetime by employing the standard
perturbative RG-techniques of particle physics (see
\cite{SolaReview11,SolaReview13} for recent reviews). These RG-based
dynamical vacuum energy models emphasize on the evolution of the vacuum
energy as a particularly well-motivated function of the Hubble rate, i.e.
$\Lambda(t)=\Lambda(H(t))$, namely functions containing even powers of $H$
and including also an additive constant term. These proposals were
confronted with the first supernovae data in\,\cite{ShapSol03}, and later
on with the modern observations on supernovae, baryonic acoustic
oscillations, CMB and structure formation
in\,\cite{BPS09,GrandeET11,BPS12,FSS06,Mirage2013}. Variants of these
models facing efficiently the cosmic coincidence problem and some aspects
of the CC problem also exist in the
literature\,\cite{LXCDM,RelaxedUniverse}, including the implications on
the possible variability of the fundamental
constants\,\cite{FritzschSola2012}.
{As remarked before, there is an extensive (old and new) literature in
which the time-evolving vacuum has been phenomenologically modeled as a
function of time in various possible ways, in particular, as a function of
the Hubble parameter
\cite{CarvalhoET92,LM93,Waga93,SalimWaga93,LM94,LT96,ML2000,ArcuriWaga94,Arbab97,OverduinCooper98,
BertolamiMartins00,Cunha02,Wang:2004cp,OpherPelinson04,BarrowClifton06,MontenegroCarneiro07,Bauer05,CT06,Basilakos09,LBS12,BLS13}.}

Technically speaking, it would be important if we could find a way to
unify all the stages of the history of the universe within the generic
framework of the running vacuum models, as these are the closest ones to
fundamental QFT physics. While a first formulation of this unification was
given in \cite{LBS12,BLS13}, the aim of the current work is to put forward
a large class of models of this kind in which the vacuum dynamics of the
early universe is linked with that of the late universe in a way fully
consistent with the phenomenological observations. At the same time we
suggest possible clues to solve or alleviate some of the fundamental
problems of the early universe, most particularly the transition from the
inflationary epoch to the standard radiation epoch.
It starts from a nonsingular inflationary stage which has a natural
(universal) ending into the radiation phase (thereby alleviating the
horizon and graceful exit problems), and, finally, the small current value
of the vacuum energy density can be conceived as a result of the massive
disintegration of the vacuum into matter during the primordial stages.

The plan of the paper is as follows. In Sec. II we discuss the energy
conservation in general dynamical models of the vacuum energy, whereas in
Secs. III and IV we motivate in different ways the form $\CC=\CC(H)$ we
are interested in. In Secs. V to VII we provide the analytical solutions
in the early and late universe respectively (the formulation in terms of
an effective potential is presented in Sec. VI). The summary and general
discussion is provided in Sec. VIII. Finally, in the appendix we furnish
some additional technical details related to the derivation of the
cosmological equations for the models under consideration.

\section{Models with dynamical vacuum energy}

In the current article we would like to investigate the cosmic expansion
within the context of the time varying vacuum energy density. To start
with, let us model the expanding universe as a mixture of perfect fluids
$N=1,2,..$ with 4-velocity fields $U_\mu^N$ and total energy momentum
tensor given by
\begin{equation}\label{Tmunu}
{T}_{\mu\nu} = \sum_N {T}^N_{\mu\nu}=\sum_N\left[ -
p_N\,g_{\mu\nu}+\big(\rho_N + p_N\,\big)\,U_{\mu}^N\,U_{\nu}^N\right]\,.
\end{equation}
The components of $T^{\mu}_{\nu}$ are the following:
\begin{equation}\label{Tcomp}
{T}^0_{0}=\sum_N \rho_N \equiv \rho_{\rm T} \,,\ \
{T}^i_j=-\sum_N p_N \,\delta^i_j\equiv -p_{\rm T}\,\delta^i_j \,,
\end{equation}
where $ \rho_{\rm T}$ and $p_{\rm T}$ are the total energy density and
pressure in the comoving frame $(U_N^0,U_N^i)=(1,0)$, respectively.
Consider now the covariant local conservation law for the mixture,
$\nabla_\mu {T}^{\mu\nu} = 0$. This expression can be worked out
explicitly from (\ref{Tmunu}), and then we can contract the result with
$U_{\nu}^N$ and use the relation $U_{\nu}^N\nabla_{\mu}U^{\nu}_N=0$ (which
follows immediately from the fact that for any four-velocity vector, we
have $U^{\mu}_N\,U_{\mu}^N=1$). The final result reads\,\cite{GPS08}
\begin{equation}\label{conserv3}
\sum_N\left[U_N^\mu\,\nabla_{\mu}\,\rho_N+(\rho_N+p_N)\nabla_\mu
{U}^\mu_N\right]=0\,.
\end{equation}
This equation is the local conservation law in a more explicit form, but
we can still further reduce it. {In the case of a
Friedmann-Lema\^{\i}tre-Robertson-Walker (FLRW) metric, it is
straightforward to check that for a comoving frame (${U}^\mu_N =
\delta^{\mu}_{0}$), one finds:}
\begin{equation}\label{nablaU3H}
\nabla_\mu {U}^\mu_N=3\,H\,\ \ \ (N=1,2,...)\,,
\end{equation}
and this relation implies that
Eq.(\ref{conserv3}) boils down to
\begin{equation}\label{EnergyCons}
\sum_N\left[\, \dot{\rho}_N + 3 H (\rho_N+p_N)\,\right]
=0\,.
\end{equation}
This is the overall conservation law of the fluid mixture in its final and
useful form\,\cite{GPS08}.

Up to this point we did not specify the nature of the fluids involved. Let
us now assume that we have a mixture of two fluids, matter and vacuum
energy. The matter fluid itself is in general a mixture of relativistic
matter (i.e. radiation, $\rho_r$) and nonrelativistic matter (i.e. cold
matter, $\rho_m$) components, {but for simplicity we address here a
situation in which there is a single matter component that dominates.}
This component can either be $\rho_r$ (in the early universe after
inflation) or $\rho_m$ (well after equality). However, when we discuss a
generic epoch we shall denote by $\rho$ the density for the (dominant)
matter component or $\omega$-fluid, whatever it be (radiation or cold
matter) and by $\rho_{\CC}$ the vacuum energy density, where
$\rho_{\CC}=\CC/(8\pi G$) in natural units. The corresponding pressures
for matter and vacuum energy are indicated by $P$ and $P_{\CC}$,
respectively. The equations of state of the two fluids are: $P=\omega\rho$
and $P_{\CC}=-\rL$ (i.e. $\omega_{\CC}=-1$), where the equation of state
(EoS) parameter for the $\omega$-fluid is a positive constant for a
spatially flat FLRW metric. In our case, $\omega=1/3$ for dominant
relativistic matter (i.e. when $\rho=\rho_r$) and $\omega=0$ for dominant
cold matter ($\rho=\rho_m$). The corresponding Einstein field equations of
the system formed by a dominant matter component and the vacuum fluid read
\begin{eqnarray}
&&8\pi G\rho_{\rm T}\equiv 8\pi G\rho + \Lambda=3H^2 \label{EE}\\
&&8\pi G p_{\rm T} \equiv 8\pi G P-\Lambda=-2\dot H-3H^2\label{EE2}\,,
\end{eqnarray}
where $H\equiv\dot a/a$ is the Hubble rate, $a=a(t)$ is the scale factor,
and the overdot denotes derivative with respect to the cosmic time $t$.
Let us note that if we consider the two Eqs.\,(\ref{EE})-(\ref{EE2})
together with the overall conservation law (\ref{EnergyCons}), only two of
them are independent. For example, if we take  the above pair as the two
independent equations, then one can easily show that (\ref{EnergyCons}) is
just a first integral of the system. However, for convenience we may also
be interested in using, say, Eq.\,(\ref{EE}) and the overall conservation
law (\ref{EnergyCons}). These two are also independent. It should then be
clear that any two of the three equations contain all the information and
the third one is identically satisfied.

Let us now discuss the possibility, in contrast to $\Lambda$CDM case, that
$\Lambda$ is not constant but a function of the cosmic time, i.e.
$\rL=\rL(t)$. This is perfectly allowed by the cosmological principle
embodied in the FLRW metric. The EoS for the vacuum and matter fluids can
still be $P_\Lambda(t)=-\rho_\Lambda(t)$ and ${P}/\rho=\omega$,
respectively, where the latter takes the aforementioned values in the
relativistic and nonrelativistic regimes. It is important to realize that
under these conditions the above Eqs.\, (\ref{EnergyCons})-(\ref{EE2})
stay formally the same, as it is easy to check. Therefore, applying the
conservation law (\ref{EnergyCons}) for a dominant matter $\omega$-fluid
plus a time-evolving vacuum ($\omega_{\CC}=-1$), we find:
\begin{equation}
\label{lambdavar}
\dot\rho_\Lambda+\dot\rho+3(1+\omega)\rho\,H=0\,.
\end{equation}
This law is a consequence of imposing the covariant conservation of the
total energy density of the combined system of matter and vacuum, and
therefore is a direct reflection of the Bianchi identity satisfied by the
geometric side of the Einstein's equations. Such law will play an
important role in our discussions. In the $\CC$CDM model, where
$\rL=$const., it is obvious that it boils down to the standard matter
conservation law $\dot\rho+3(1+\omega)\rho\,H=0$.

\section{General ansatz for the evolving vacuum as a function of
$H$}\label{sect:GeneralAnsatz}

Our main aim in this paper is to study a relevant class of time-evolving
models for the vacuum energy. However, we do not aim at an arbitrary
function of the cosmic time $\CC=\CC(t)$. In fact, we focus on a dynamical
CC term, $\CC$, whose primary dependence is on the Hubble rate and from
here the vacuum energy inherits its time dependence: $\CC(t)=\CC(H(t))$.
As we will see, this is more in consonance with the expectations in QFT.
Nonetheless not all possible functional dependences on $H$ are allowed. In
order to obtain a definite decaying $\Lambda$ cosmology we need to find a
viable expression for $\Lambda$ in terms of the Hubble rate. The
motivation for a function $\CC=\CC(H)$ can be provided from different
points of view. Let us start from a general phenomenological one, and only
afterwards (see the next section) we will motivate it in more formal
terms. The existence of two fluid components means that we may introduce
the following ratio:

\begin{equation}\label{eq:ratio}
\beta(t) =  \frac{\rL - \rho_{\Lambda 0}}{\rho + \rL},
\end{equation}
where $\rho_{\Lambda 0}$ is a constant vacuum density defining the
fiducial constant $\Lambda$. This $\beta (t)$  parameter quantifies the
time variation of the vacuum energy density. It has the
following properties:\\
(i) If $\rL=\rho_{\Lambda 0}$, then $\beta=0$, and the model is
$\Lambda$CDM, (ii) If $\rho_{\Lambda 0}= 0$, then the ratio
(\ref{eq:ratio}) defines a fraction of the vacuum to the total density. If
this fraction is constant in the course of the cosmic evolution we have:
\begin{equation}
\rL = \beta \rho_{\rm T},
\end{equation}
or, equivalently, from  Eq.\,(\ref{EE}),
\begin{equation}\label{eq:bH2}
\Lambda = 3\beta H^{2}.
\end{equation}
This kind of model was discussed long ago by many authors
\cite{FreeseET87,CarvalhoET92,ArcuriWaga94}. It needs only the assumption
that the ratio (\ref{eq:ratio}) remains constant. However, when confronted
with the current observations it provides a poor fit\,\cite{BPS09}. As a
matter of fact, it is ruled out by an even more fundamental reason,
because in these models there does not exist a transition redshift from
deceleration to acceleration as required by supernovae data. {The ansatz
(\ref{eq:bH2}) implies that the universe is always accelerating or
decelerating depending} on the value of $\beta$. A brief discussion on
this point is presented at the end of Sec. \ref{sect:MatterEraDE}, see
also\,\cite{BPS12} for a more detailed discussion. More recently, this
$\Lambda(H)$-law has also been applied to discuss the late stages of the
gravitational collapse \cite{CL12}.

If we, instead, consider that the ratio given by (\ref{eq:ratio}) is
constant, then we have:
\begin{equation}\label{eq:RGoriginal}
\Lambda(t) = c_0 + 3\beta H^{2}(t)\,,
\end{equation}
where $c_0= 8\pi G \rho_{\Lambda 0}$. Notice that the present value of the
CC in this framework reads $\CC_0=c_0+3\beta H_0^2$. Such a model was
first proposed in \cite{ShapSol00} from the point of view of the RG and it
has been studied extensively in the literature, cf. Refs.
\cite{ShapSol03,Wang:2004cp,BPS09,GrandeET11,BPS12}. In contrast to
(\ref{eq:bH2}) the presence of the additive term is well-motivated within
the RG approach (see Sect. \ref{sect:runningV} ) and allows the existence
of a transition from deceleration to acceleration, and of course then also
a smooth connection with the $\CC$CDM model is possible in the limit
$\beta\to 0$. Notice that, in contrast, the model (\ref{eq:bH2}) has no
$\CC$CDM limit. In general the ratio (\ref{eq:ratio}) may not remain
constant during the evolution i.e. $\beta$ should be a time-dependent
quantity. In this case the vacuum energy density reads
\begin{equation}
\rL = \rho_{\Lambda 0}  + \beta(t) \rho_{\rm T},
\end{equation}
or equivalently
\begin{equation}\label{eq:betat}
\Lambda(t) = c_0 + 3\beta (t) H^{2}.
\end{equation}
Since $\beta(t)$ is now variable, the value of the current CC is
$\CC_0=c_0+3\beta(t_0) H_0^2$, where $t_0$ is the present cosmic time. Let
us assume that we can expand the time-dependent parameter $\beta(t)$ as
follows: $\beta(t) = \nu + \alpha(\frac{H}{H_I})^{n}$, where $\nu$,
$\alpha$ and $H_I$ are constants whose interpretation will become apparent
later on, and $n$ is typically a positive integer $n\geqslant 1$. The
expansion of $\beta(t)$ in this form can be seen as a constant term plus a
time-dependent term.  The latter should naturally depend on a power of the
expansion rate, $n=1$ being the {simplest} possibility (although other
constraints could change this option). Several aspects of the  case $n=1$
with a flat geometry were discussed long ago
in\,\cite{LM94}, and, later on, the case for closed and hyperbolic
geometries was also investigated \cite{LT96}. The case with $c_0=0$ and
$\nu = 1-\beta$ and arbitrary values of $n$ was first phenomenologically
proposed in \cite{ML2000} while the case $n=2$ with $c_0=\beta=0$ (plus a
linear term in $H$) was more recently investigated in \cite{CT06}. In
general, for the above $\beta(t)$  we arrive at the general ansatz:
\begin{equation}\label{lambda}
\Lambda(H) = c_0 + 3\nu H^{2} + 3\alpha
\frac{H^{k}}{H_{I}^{(k-2)}} \;,
\end{equation}
where $k=n+2$. It is interesting to note that the next-to-leading
higher order power, i.e. the case $k=4$, can be motivated on more
fundamental QFT grounds, as shown long ago in\,\cite{ShapSol01} and more
recently in \cite{Fossil07} within the framework of the modified
anomaly-induced inflation scenarios. These are a generalization of
Starobinsky's model type of inflation\,\cite{Starobinsky80}, in which the
vacuum effective action for massive quantum fields can be computed using
the conformal representation of the fields action\,\cite{PSW87}.

If we would not attend other considerations, the integer $k$ in
Eq.(\ref{lambda}) is generally unrestricted, apart from $k\geqslant 3$.
Obviously the case $k=2$ (i.e. $n=0$) is not considered because it
corresponds to the situation $\beta=\nu+\alpha=$const., considered in the
original RG formulation (\ref{eq:RGoriginal}) (see next section) which
already contains $H^2$ as the highest power of the Hubble rate. This
situation is equivalent to $k=0$ upon redefining $c_0$ and with
$\beta=\nu=$const. Nontrivial departure of these cases thus requires
$k\geqslant 3$.

The constant additive term in (\ref{lambda}) obviously represents the
dominant contribution at very low energies (i.e. when
$H\approx{\mathcal{O}}(H_0)\ll H_I$). The $H^2$ term represents a small
correction (if $\nu\ll1$) to the dominant term at the present time. While
it provides a mild time-evolving behavior to the vacuum energy density at
intermediate times. On the other hand, the $H^k$ ($k\geqslant 3$) power
acquires a great relevance in the early universe, near the $H_I$ energy
scale -- interpreted as the inflationary expansion rate.

Since $H_I$ is presumably large, it is clear that $\beta(t_0)\simeq\nu$
for any $n$ and hence the value of the CC today is essentially
$\CC_0=c_0+3\nu H_0^2$ for all models of the class (\ref{lambda}). Thus,
effectively, for any $k\geqslant3$ the proposed model (\ref{lambda}) is
very close to the model (\ref{eq:RGoriginal}) for a description of the
postinflationary cosmology, including of course the evolution near the
current time. It follows that the coefficient $\nu$ is the relevant one
for the dynamical evolution of the vacuum energy in most of the universe's
history. However, for the early universe the additional term $H^k$ takes
over and the effective behavior of Eq.\,(\ref{lambda}) is then
$\Lambda(t)\simeq 3\alpha {H^{k}(t)}/{H_{I}^{(k-2)}}$ ($k\geqslant3$), and
here the relevant coefficient is $\alpha$ together with the inflationary
scale $H_I$. Of course $\alpha$ and $H_I$ appear to be a convenient way to
break down the single coefficient of the dominant power $H^k$. To
disentangle the value of the dimensionless coefficient $\alpha$ we would
need to relate $H_I$ to some physical high-energy scale, for example a
typical grand unified theory (GUT) scale associated to the inflationary
time.

Let us mention that the covariance of the effective action of QFT in
curved spacetime indicates that the even powers of $H$ are preferred (see
the next section); in other words, the new term $H^k$ correcting the
original expression (\ref{eq:RGoriginal}) is expected to have $k=2m$ (with
$m=2,3,...$). {Naturally, despite the fact that the odd powers $k=3,5,...$
in (\ref{lambda}) are not favored, we will not completely neglect them, if
only from the phenomenological point of view (see Refs. \cite{LM94,LT96}).
In contrast, the case $k=0$ leads to the model (\ref{eq:RGoriginal})
considered in\,\cite{ShapSol00}, which is adequate for the more recent
universe\,\cite{ShapSol03,BPS09,GrandeET11} but not for the very early
stages. In this paper, we are proposing a generalization that leads to the
unification model (\ref{lambda}) for the complete description of the
cosmological history from the very early universe to the present time.}

Now, combining Eqs.(\ref{EE}), (\ref{lambdavar}), (\ref{lambda}), and
using the EoS of the fluid components we obtain the following
differential equation for the time evolution of the Hubble
parameter:
\begin{equation}
\label{HE}
\dot H+\frac{3}{2}(1+\omega)H^2\left[1-\nu-\frac{c_0}{3H^2}-
\alpha\left(\frac{H}{H_I}\right)^{n}\right]=0.
\end{equation}
Remarkably there are two constant value solutions to this equation, namely
$H=H_I[(1-\nu)/\alpha]^{1/n}$, corresponding to the very early universe,
i.e. when $c_0\ll H^2$. On the other hand, at late times, when $H\ll H_I$ we
have $H=[c_0/3(1-\nu)]^{1/2}$, whereby $\Lambda\approx c_0$ which behaves as an
effective cosmological constant. Also using Eq.\eqref{HE} the
deceleration parameter $q\equiv-\ddot a a/\dot a^2=-1-\dot{H}/H^2$ is
given by
\begin{equation}
\label{q}
 q(H)=\frac{3}{2}(1+\omega)
\left[1-\nu-\frac{c_0}{3H^2}-\alpha\left(\frac{H}{H_I}\right)^n\right]-1\,.
\end{equation}

We shall present below the various phases of the
decaying vacuum cosmology \eqref{lambda}, starting
from an unstable inflationary phase 
powered by the huge value $H_I$ presumably connected to the scale of a GUT
or even the Planck scale $M_P$, then it deflates (with a massive
production of relativistic particles), and subsequently evolves into the
standard radiation and matter eras. Finally, it effectively appears today
as a slowly dynamical dark energy.

\section{Running vacuum $\CC=\CC(H)$}\label{sect:runningV}
In the previous section we have motivated the time evolution of the vacuum
energy density as a function of the Hubble rate using a general
phenomenological argumentation. However, the running of the vacuum energy
is expected in QFT in curved spacetime on more fundamental
grounds\,\cite{ShapSol00,Fossil07,ShapSol09}, see also
\cite{SolaReview11,SolaReview13} and references therein. Running couplings
in flat QFT provide a useful theoretical tool to investigate theories as
QED or QCD, where the corresponding gauge coupling constants run with a
scale $\mu$ associated to the typical energy of the process, $g=g(\mu)$.
Similarly, in the effective action of QFT in curved spacetime $\rL$ should
be an effective coupling depending on a mass scale $\mu$. In the universe
we should expect that the running of $\rL$ from the quantum effects of the
matter fields is associated with the change of the spacetime curvature,
and hence with the change of the typical energy of the classical
gravitational external field linked to the FLRW metric. As this energy is
pumped into the matter loops from the tails of the external gravitational
field, it could be responsible for the physical running. Therefore we
naturally associate $\mu^2$ to $R$, where (for flat FLRW metric)
\begin{equation}\label{eq:Rset}
|R|=6\left(\,\frac{\ddot{a}}{a}+\frac{\dot{a}^2}{a^2}\,\right)
=12\,H^2+6\,\dot{H}\,.
\end{equation}
It follows that $\mu^2$ is in correspondence with  $H^2$ and $\dot{H}$.
For simplicity we concentrate on the setting $\mu=H$ as we expect that it
already captures the essential dynamics (see \cite{BPS12}). Within this RG
approach the rate of change of $\rL$ with $\mu=H$ should satisfy a
corresponding RG equation:
\begin{eqnarray}\label{eq:seriesRL}
(4\pi)^2\frac{d\rL(\mu)}{d\ln\mu^2}&=&
\sum_{m=1,2,...}\,A_{2m}\,\mu^{2m}\nonumber\\
&=&A_2\,\mu^2+A_4\,\mu^4+A_6\,\mu^6...\nonumber\\ \label{seriesRGG}
\end{eqnarray}
The \textit{r.h.s.} of this expression defines essentially the
$\beta$-function for the RG running of $\rL$. The coefficients $A_{2m}$
receive loop contributions from boson and fermion matter fields of
different masses $M_i$. Notice that only the even powers of $\mu=H$ are
involved, since in this formulation $\rL(H)$ is of course part of the
effective action of QFT in curved spacetime and hence it should be a
covariant quantity\,\cite{ShapSol00,Fossil07,ShapSol09}. Worth noting is
that we have omitted the $A_0$ term in (\ref{eq:seriesRL}), as it would be
of order $M_i^4$ and hence would trigger a too fast running of $\rL$. This
can also be formally justified from the fact that all known particles
satisfy $\mu<M_i$ (for $\mu=H$). Thus, since none of them is an active
degree of freedom for the running of $\rL$, only the subleading terms are
available. The first subleading term is the $A_2\mu^2$ one, where $A_2$
has dimension of mass squared, namely $A_2\sim \sum_i a_i M_i^2$ where the
sum is over the masses of all fields involved in the computation of the
$\beta$-function (including their multiplicities). Similarly, since all
the coefficients $A_{2m}$ (except $A_4$) are dimensionful it is convenient
to rewrite them appropriately in a way such that the mass dimensions are
explicit. Thus we rewrite (\ref{eq:seriesRL}) as follows:
\begin{eqnarray}\label{seriesRLH}
\frac{d\rL(\mu)}{d\ln
H^2}=\frac{1}{(4\pi)^2}\sum_{i}\left[\,a_{i}M_{i}^{2}\,H^{2}
+\,b_{i}\,H^{4}+c_{i}\frac{H^{6}}{M_{i}^{2}}\,+...\right]
\nonumber\\
\end{eqnarray}
The sum over the masses of the fields involved in the loop contributions
is now explicit. Specific realizations of the structure (\ref{seriesRLH})
can be obtained in one-loop calculations within particular frameworks, see
e.g. \,\cite{Fossil07}. As we can see, the series became now an expansion
in powers of $H$. If we integrate Eq.\,(\ref{seriesRLH}) to obtain
$\rL(H)$, an additive term (independent of $H$) obviously appears as well.
In other words, the result for $\CC(H)=8\pi G \rL(H)$ nicely adapts to the
form (\ref{lambda}) suggested by the general argument of the previous
section, which means that the RG formulation may provide a fundamental
link of that form with QFT in curved spacetime. However, as emphasized,
only the even powers of $H$ are involved in the RG realization, owing to
the general covariance of the effective action. As it is obvious, the
expansion (\ref{seriesRLH}) converges very fast at low energies, where $H$
is rather small -- certainly much smaller than any particle mass. No other
$H^{2m}$-term beyond $H^2$ (not even $H^4$) can contribute significantly
on the \textit{r.h.s.} of Eq.\,(\ref{seriesRLH}) at any stage of the
cosmological history below the GUT scale $M_X$, typically a few orders of
magnitude below the Planck scale $M_P\sim 10^{19}$ GeV.

However, if we wish to have access to the physics of inflation and in
general to the very early states of the cosmic evolution, we need to keep
at least the terms $H^4$. It is interesting to note the structure of the
leading term in the series (\ref{seriesRLH}), i.e.  $\sim \sum_i
M_i^2H^2$. This term is of course dominated by the loop contributions of
the heaviest fields with masses $M_i$ of order of $M_X$, the GUT scale
near the Planck mass. It follows that in the early universe (when $H$ is
also close, but below, $M_i\sim M_X$) the $H^4$ effects can also be
significant, whereas the terms $H^6/M_i^2$ and above are less and less
important. Therefore, the dominant part of the series (\ref{seriesRLH}) is
expected to be naturally truncated at the $H^4$ term. These terms should
contain the bulk of the high energy contributions within QFT in curved
spacetime, namely within a semiclassical description of gravity near but
(possibly a few orders) below the Planck scale. Models of inflation based
on higher order terms inspired by the RG framework have existed for a long
time in the literature, see \cite{ShapSol01} as well as the unified
inflation-dark energy framework of \cite{Fossil07} (see also
\cite{LM94,LT96,ML2000,CT06} for a more phenomenological treatment).

We can find the explicit relation between the one-loop coefficients of the
RG equation (\ref{seriesRLH}) with the phenomenological coefficients
introduced in Sect.\,\ref{sect:GeneralAnsatz}. Let us consider the case
$n=2$, for which the highest power of the Hubble rate in the vacuum
expression is $H^4$. Upon integrating the RG equation
(\ref{seriesRLH}) and comparing with Eq.\,(\ref{lambda}) we obtain
\begin{equation}\label{eq:nuloopcoeff}
\nu=\frac{1}{6\pi}\, \sum_{i=f,b} c_i\frac{M_i^2}{M_P^2}\,,
\end{equation}
and
\begin{equation}\label{eq:alphaloopcoeff}
\alpha=\frac{1}{12\pi}\, \frac{H_I^2}{M_P^2}\sum_{i=f,b} b_i\,.
\end{equation}
A few words will help to better interpret this result. First of all let us
note that $\nu$ acts indeed as the reduced (dimensionless)
$\beta$-function for the RG running of $\rL$ at low energies, whereas
$\alpha$ plays a similar role at high energies. Moreover, both
coefficients are predicted to be naturally small because $M_i^2\ll M_P^2$
for all the particles, even for the heavy fields of a typical GUT. In the
case of the low energy coefficient $\nu$ a concrete realization of the
structure (\ref{eq:nuloopcoeff}) is given in \cite{Fossil07},  and an
estimate within a generic GUT is found in the range
$|\nu|=10^{-6}-10^{-3}$. Similarly, the dimensionless coefficient $\alpha$
is naturally predicted small, $|\alpha|\ll 1$, because the inflationary
scale $H_I$ is certainly below the Planck scale $M_P$. In a typical GUT
where $M_X\sim 10^{16}$ GeV$^4$ we have  $H_I/M_P\sim M_X^2/M_P^2\lesssim
10^{-6}$. Even counting the large multiplicities of the fields in usual
GUT's, the two coefficients $\nu$ and $\alpha$ are expected to be rather
small, which is indeed the natural expectation since they play the role of
one-loop $\beta$-functions at the respective low and high energy scales.
Using a joint likelihood analysis of the recent supernovae type Ia data,
the CMB shift parameter, and the baryonic acoustic oscillations one finds
that the best fit value for $\nu$ in the case of a flat universe is at
most of order $|\nu|={\cal O}(10^{-3})$\,\cite{BPS09,GrandeET11}, which is
nicely in accordance with the aforementioned theoretical expectations.

\section{From the early de Sitter stage to the radiation
phase}\label{sect:deSitterRadiation}

\begin{figure*}\label{fig1}
\centerline{\epsfig{file=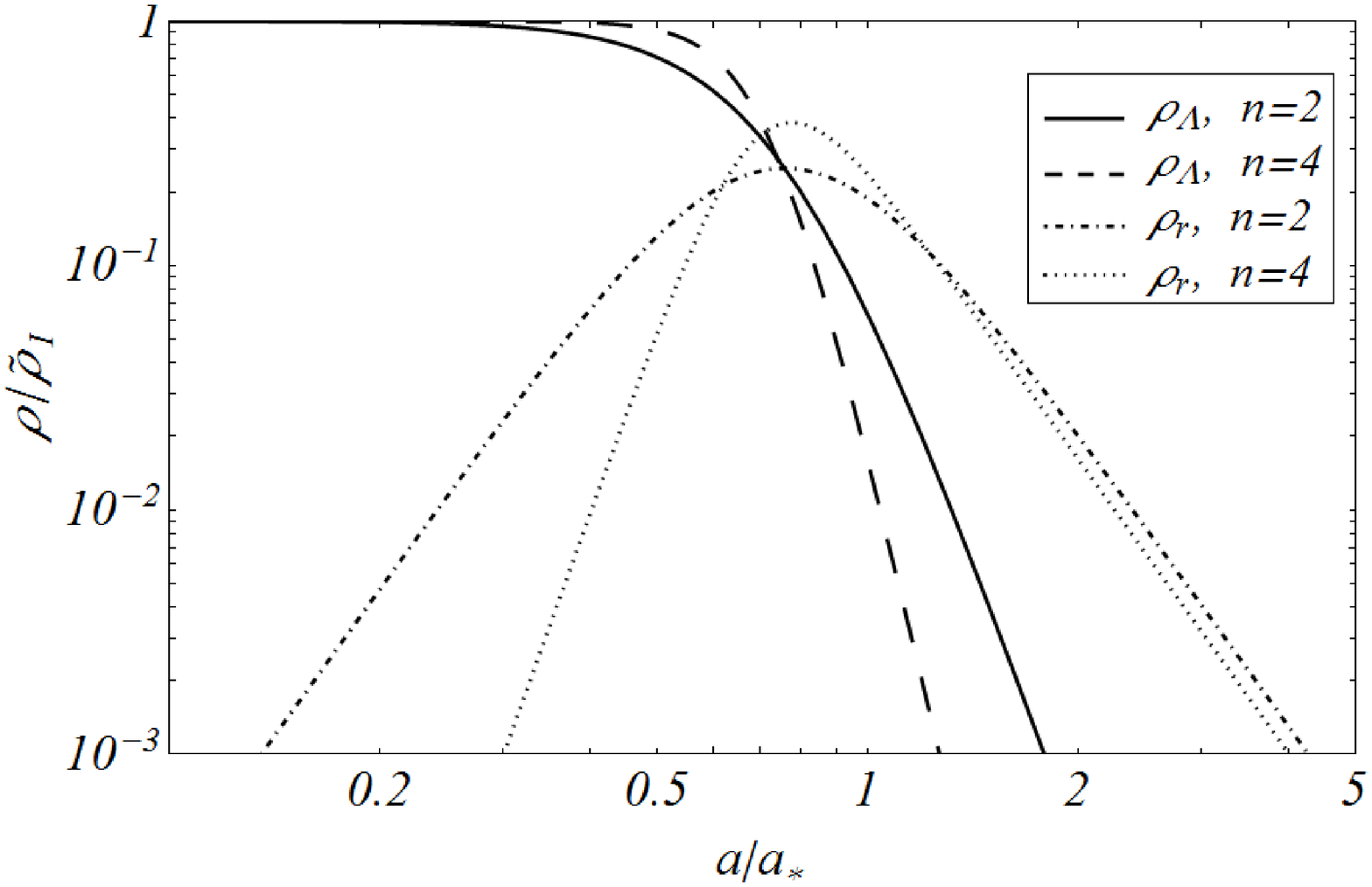,width=8.5cm}\hspace{0.5cm}\epsfig{file=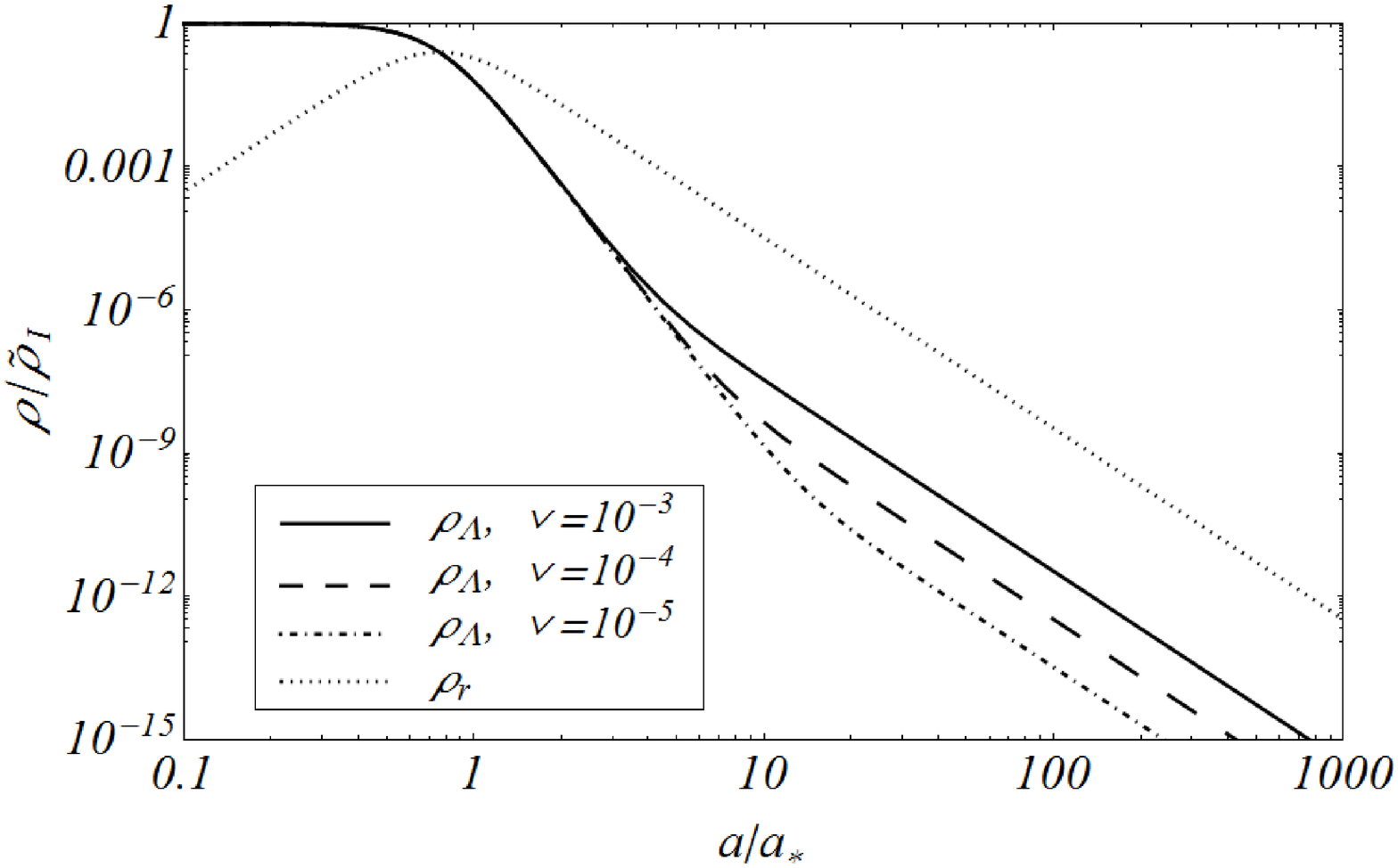,width=8.7cm}
 \hskip 0.1in}
\caption{{\em Left panel}: The evolution of the vacuum and radiation
energy densities during the primordial era, where $H^2\gg c_0$. We
normalize the densities with respect to the primeval critical value
$\tilde{\rho}_I$ defined in (\ref{eq:rhoItild}). The plots show that the
decay of the vacuum density, as well as the production and subsequently
dilution of radiation, occur in a faster way for large values of the
parameter $n$ [recall that $k=n+2$ in Eq.\,(\ref{lambda})], thereby
ensuring the universality of the graceful exit for any $n\geq2$.
Additionally, in this figure we can see that the vacuum density always
decays faster than it does the radiation density after the transition
period. {\em Right panel}: The behavior of the vacuum density with the
variation of the parameter $\nu$ for $n=2$. In this graph, we can see that
during the radiation dominated era the vacuum density ceases to decay; it
only dilutes with time (in a similar way as the radiation energy density)
due to the effect of the expansion. The precise instant when this change
occurs is earlier for larger values of the parameter $\nu$. On the other
hand, the evolution of the radiation energy density is affected very
little by the variation of the parameter $\nu$, for $\nu\leq10^{-3}$. In
this figure we show the radiation energy density for $\nu=10^{-4}$.}
\end{figure*}

Let us first discuss the transition from the initial de Sitter stage to
the radiation phase, while $c_0\ll H^2$. The solution \eqref{HS1a} -- see
Appendix -- of Eq.\,\eqref{HE} for $\omega=1/3$ and $c_0=0$ becomes
\begin{equation}\label{HS1}
 H(a)=\frac{\tilde H_I}{\left[1+D\,a^{2\,n\,(1-\nu)}\right]^{1/n}}\,,
\end{equation}
where we have defined $\tilde H_I \equiv \left(\frac{1-\nu}{\alpha}\right)^{1/n} H_I$, is the critical
Hubble parameter associated to the initial de Sitter era,
or
\begin{equation}\label{HS1b}
\int_{a_\star}^a\frac{d\tilde{a}}{\tilde{a}}\left[1+D\,\tilde{a}^{2\,n\,(1-\nu)}\right]^{1/n}= \tilde H_It
\end{equation}
where $t$ here is the time elapsed after (approximately) the end of the
inflationary period, indicated by $t_\star$, and we have defined
$a_\star=a(t_\star)$. The integration constant $D$ is fixed from the
condition $H(a_\star)\equiv H_\star$, thus
\begin{equation}\label{eq:defD}
D=a_\star^{-2\,n\,(1-\nu)}\left[\left(\frac{\tilde H_I}{H_\star}\right)^n-1\right]\,.\end{equation}

Equation (\ref{HS1b}) will be useful below for particular considerations.
However, rather than directly integrating this equation it is possible to
retake (\ref{HS1}) and cast it in a more appropriate form that allows to
express the result $t=t(a)$ in terms of special functions. This is done in
the Appendix. The final result is
\begin{equation}\label{tn}
\begin{split}
t(a)=&\frac{\left(1+D\,a^{2n(1-\nu)}\right)^{\frac{1+n}{n}}}{2(1-\nu)\tilde H_I\,D\, a^{2n(1-\nu)}} \times\\
       & F\left[1\,,1\,,1-\frac{1}{n}\,,\frac{-1}{D\,a^{2n(1-\nu)}}\right]\,,
\end{split}
\end{equation}
where $F[\alpha_1,\alpha_2,\alpha_3,z]$ is the Gauss hypergeometric
function, and as in (\ref{HS1b})  we count the time passed after
$t_{\star}$, i.e. $t$ is the cosmic time within the FLRW regime. Using the
Einstein equations and the above solutions we can obtain the corresponding
vacuum, radiation and total energy densities:
\begin{equation}\label{eq:rLa}
  \rho_\Lambda(a)=\tilde{\rho}_I\,\frac{1+\nu\,D\,a^{2n(1-\nu)}}{\left[1+D\,a^{2n(1-\nu)}\right]^{1+2/n}}\,,
\end{equation}
\begin{equation}\label{rho_1}
 \rho_r(a)=\tilde{\rho}_I\,\frac{(1-\nu)D\,a^{2n(1-\nu)}}{\left[1+D\,a^{2n(1-\nu)}\right]^{1+2/n}}\,,
\end{equation}
\begin{equation}
 \rho_{\rm T}(a)=\tilde{\rho}_I\,\frac{1}{\left[1+D\,a^{2n(1-\nu)}\right]^{2/n}}\,.
\end{equation}
where we have defined
\begin{equation}\label{eq:rhoItild}
\tilde{\rho}_I\equiv\frac{3\tilde H_I^2}{8\pi G}
\end{equation}
is the primeval critical energy density associated with the initial de
Sitter stage. We can see from (\ref{eq:rLa}) that the value
(\ref{eq:rhoItild}) just provides the vacuum energy density for $a\to 0$,
namely $\rho_{\CC}(0)=\tilde{\rho}_I$. As $|\nu|\ll 1$ we have
$\tilde{\rho}_I /\rho_I\sim\alpha^{-2/n}$ and hence the density
$\tilde{\rho}_I $ can differ a few orders of magnitude from $\rho_I$ since
we also expect (see the previous section) that $|\alpha|\ll 1$. Let us
also emphasize from the previous formulas that for $a\to 0$ we have
$\rho_r/\rL\propto a^{2n(1-\nu)}\to 0$, i.e. the very early universe is
indeed vacuum-dominated with a negligible amount of radiation. {For the
numerical analysis} of the energy densities, see {Fig.\,1}.

Notice that the constant (\ref{eq:defD}) entering Eq.\,(\ref{HS1}) is
greater than zero precisely for $\tilde H_I>H_\star$, which
is tantamount to say $\rho_{\star}<\tilde{\rho}_I$, where
$\rho_{\star}\equiv 3H_{\star}^2/8\pi G$ is the critical energy density at
the time $t=t_{\star}$. The existence of a point marking the decrease of
the energy density from the initial steady value $\tilde{\rho}_I$ is
indeed the condition that points to a deflationary period after inflation.

\begin{figure*}\label{fig3}
\centerline{\epsfig{file=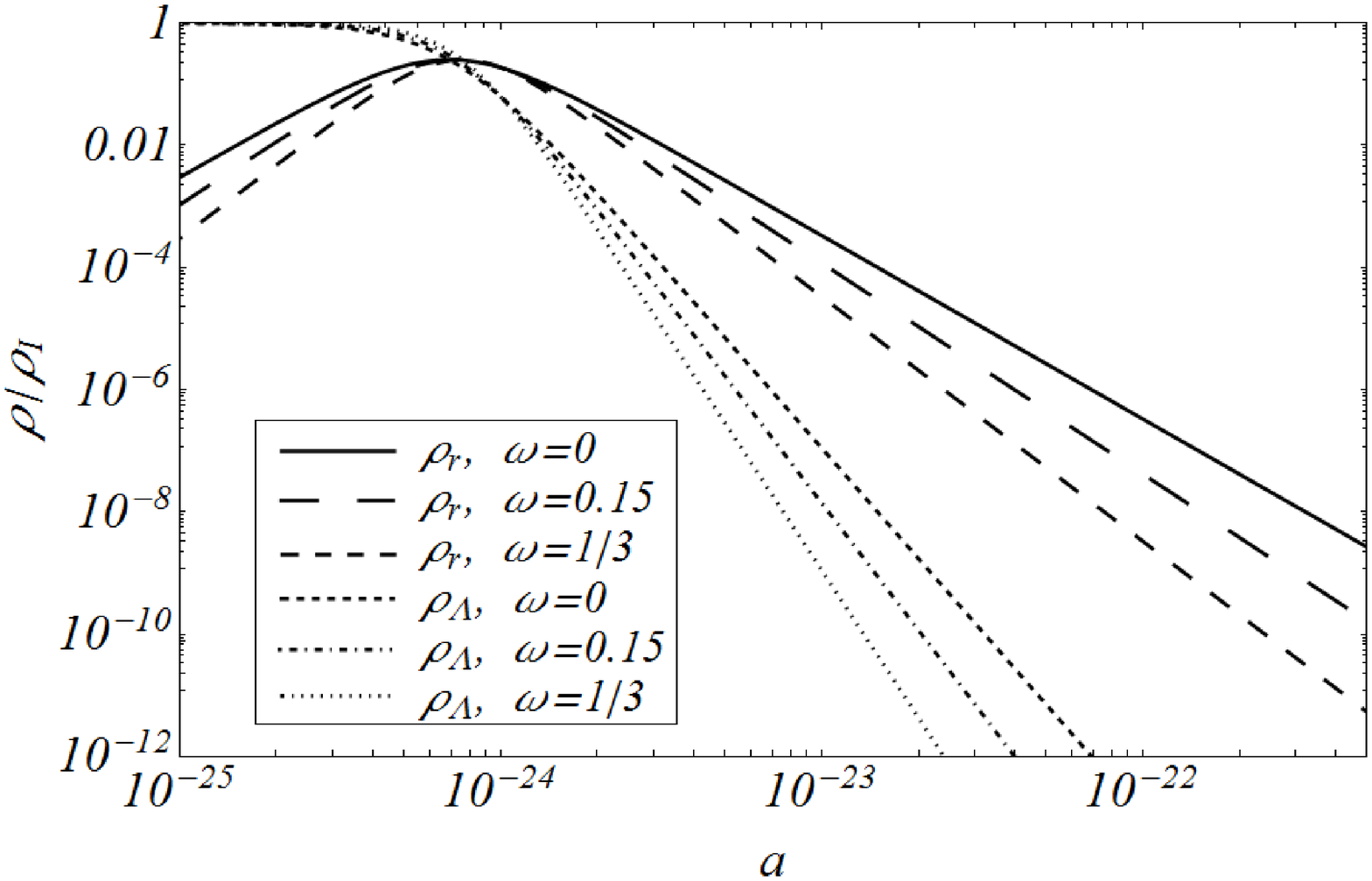,width=8.5cm,height=6cm}\hspace{0.5cm}\epsfig{file=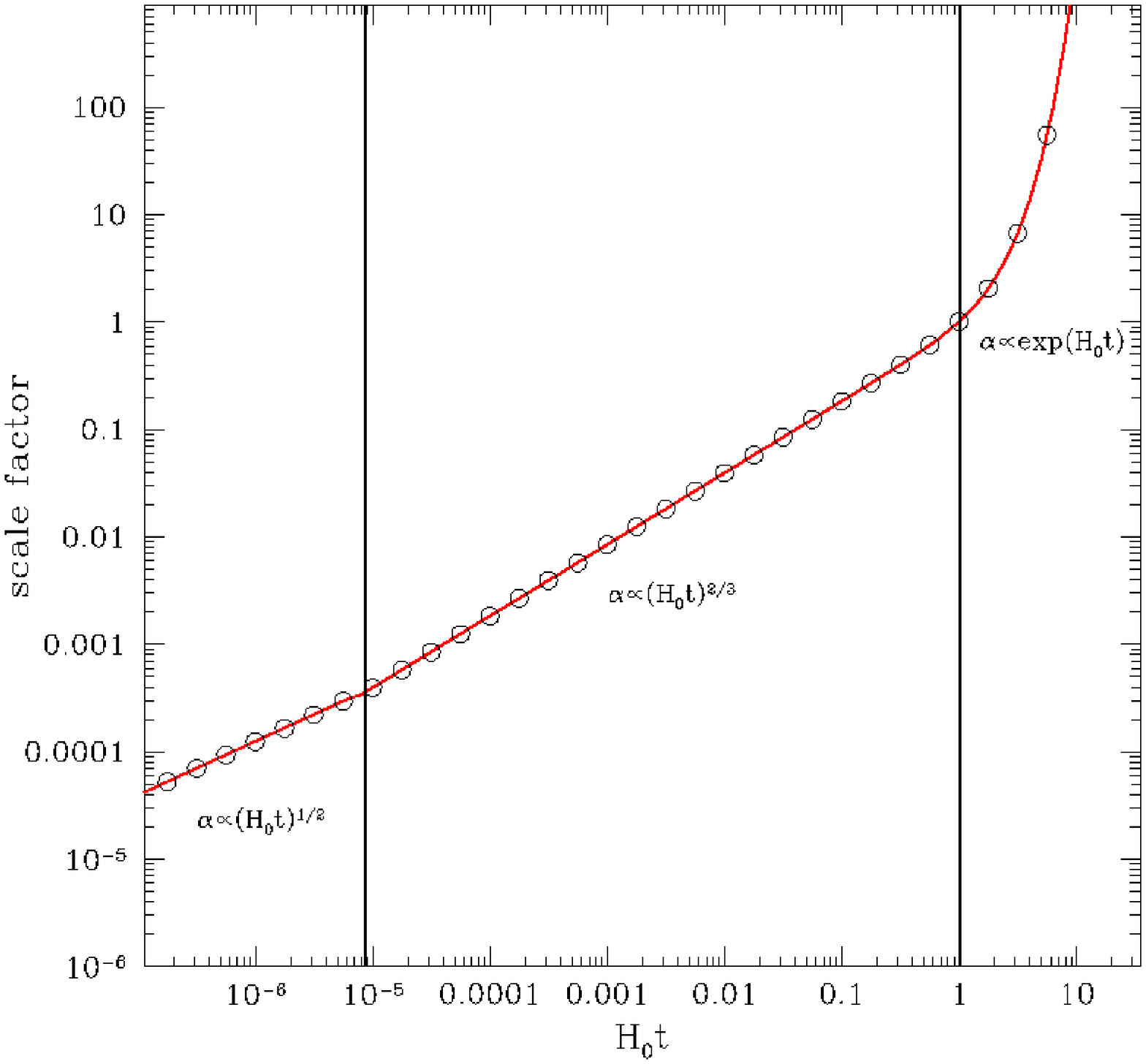,width=8.5cm,height=6.2cm}
 \hskip 0.1in}
\caption{{\em Left panel}: Universality of the graceful exit with respect
to the variation of the EoS parameter for the matter $\omega$-fluid . Once
more we normalize the densities with respect to the primeval critical
value $\tilde{\rho}_I$ defined in (\ref{eq:rhoItild}). This figure shows
that the rate of the vacuum decay, radiation production and subsequent
dilution is larger for greater values of $\omega$. As always, the vacuum
decay is faster than the rate of dilution of the radiation, thereby
ensuring the transition toward a radiation domination era after the end of
the inflationary stage. {\em Right panel}: The evolution of the scale
factor predicted by the decaying $\Lambda(H)$ model at the late stage
$H_{I} \gg H$ for $\nu \sim 10^{-3}$ (solid line) versus the traditional
$\Lambda$CDM cosmology (open points). In this plot we have adopted the
best fit, ${\Omega}_{\Lambda}^{0}=0.6825$, from the recent results of
PLANCK data \cite{Ade13}. Clearly, the expansion history of the scale
factor of the $\Lambda(H)$ model is almost indistinguishable from the
$\CC$CDM model for the entire postinflationary era up to our days, and
into the future. }
\end{figure*}

For $D\,a^{2n(1-\nu)}\ll1$ (during the very early universe) the solution
\eqref{HS1} can be approximated by the constant value solution $H\approx
\tilde H_I$. As mentioned, the vacuum energy density
remains almost constant $\rho_\Lambda\approx \tilde{\rho}_I$ in this
period and coexists with a negligible radiation density, which just starts
to grow as $\rho_r\simeq \rho_\Lambda(1-\nu)D\,a^{2n(1-\nu)}$. This stage
obviously depicts the primeval de Sitter era in the cosmic evolution, with
\begin{equation}\label{eq:deSitter1}
a(t)\propto \exp[\left\{\tilde H_I t\right\}]\,,
\end{equation}
in which the universe undergoes a process of primordial inflation. The
result (\ref{eq:deSitter1}) can be derived by expanding the solution
\eqref{tn} around $D\,a^{2n(1-\nu)}\ll1$:
\begin{eqnarray}
\tilde H_I t\approx  \frac{1}{2n(1-\nu)} \times\left[C+\ln D\,a^{2n(1-\nu)}\right]\,,
\end{eqnarray}
where $C$ is a constant (dependent on $n$) not playing a role in this
argument. Notice that Eq.\,(\ref{eq:deSitter1}) can also be substantiated
by simply letting $a\to 0$ before integrating Eq.\,(\ref{HS1b}).

The outcome of the above considerations is that for $D\neq0$ the universe
starts without a singularity and thus this model overcomes the horizon
problem. The universe then evolves naturally toward a radiation-dominated
universe (hence providing a useful clue to explaining the ``graceful
exit'' from the inflationary stage, {see Fig. 2}).
On the other hand, a light pulse beginning at $t=-\infty$ will have
traveled by the cosmic time $t$ a physical distance, $d_{H}(t)=
a(t)\int_{-\infty}^{t}\frac{d t'}{a(t')}$, which diverges thereby implying
the absence of particle horizons, thus the local interactions may
homogenize the whole universe.

The solution for the radiation energy density \eqref{rho_1} reaches a
maximum value when the scale factor $a$ takes on the value
$a_{*}\equiv\left(2D/n\right)^{-1/2n(1-\nu)}$, which is the value when the
inflation period is accomplished and the radiation-dominated era begins
(see {Figs. 1 and 2}).

For $D\,a^{2n(1-\nu)}\gg1$ the solution \eqref{HS1}  can be approximated
as
\begin{equation}
H\approx \tilde H_I D^{-1/n} a^{-2(1-\nu)}\,,
\end{equation}
which displays the behavior $H\sim a^{-2(1-\nu)}\sim a^{-2}$ in the limit
of small $|\nu|$. Similarly from \eqref{tn} we find
\begin{equation}
t\approx\,a^{2(1-\nu)}\,.
\end{equation}
The derivation of the latter expression is particularly straightforward
from (\ref{HS1b}), if we use the limit $D\,a^{2n(1-\nu)}\gg1$ before
integration. As $|\nu|\ll 1$, it is obvious that we have essentially
reached the radiation domination era for which $a\propto
t^{1/2(1-\nu)}\simeq t^{1/2}$. This is confirmed after inspecting the
radiation density (\ref{rho_1}), which decays as
$\rho_r\propto(1-\nu)a^{-4(1-\nu)}\sim a^{-4}$. We can also see from
(\ref{eq:rLa}) that the vacuum energy density follows a similar decay law
$\rho_\Lambda\propto\nu a^{-4(1-\nu)}$, but is suppressed by the factor
$\rL/\rho_r\propto \nu$  (with $|\nu|\ll 1$) as compared to the radiation
density. This is exactly the opposite situation to the very early period
when $D\,a^{2n(1-\nu)}\ll 1$, in which the vacuum energy density is huge
and stuck at the value $\tilde{\rho}_I$ whereas the radiation density is
largely suppressed by the power $a^{2n(1-\nu)}$ of the very small scale
factor at that time. In between these two eras, we see that we can have
either huge relativistic particle production $\rho_r\propto a^{2n(1-\nu)}$
in the deflation period (namely around $D a^{2n(1-\nu)}<1$) or standard
dilution $\rho_r\propto a^{-4}$ (up to small corrections of order
$|\nu|\ll 1$) well in the radiation era ($Da^{2n(1-\nu)}\gg1$).

\subsection*{Radiation temperature}

In the case of $\omega=1/3$, we have relativistic matter production
$\rho_r\propto a^{2n(1-\nu)}$ during the deflationary era and
corresponding dilution $\rho_r\propto a^{-4(1-\nu)}$ during the subsequent
radiation-dominated era (due to the expansion of the universe).
Considering ``adiabatic'' expansion of the universe during both eras, the
radiation energy density scales as $\rho_{r} \sim T_r^4$
\cite{CalvaoET92,Lima96} with its temperature. Thus we can see that the
radiation temperature grows as $T_r\propto a^{n(1-\nu)/2}$ during the
initial de Sitter era of accelerated expansion if the specific entropy per
particle remains constant during this period. Hence, the universe is
naturally heated before it enters the radiation-dominated era.

After the de Sitter stage, the temperature decreases continuously in the
course of the expansion as $T_r\propto a^{-(1-\nu)}$, namely very close to
$1/a$ for $|\nu|\ll 1$, as it should be for an noninteracting adiabatic
expansion. Accordingly, the comoving number density of photons scales as
$n_{\gamma}\sim T^3 \propto {a}^{-3(1-\nu)}$, which shows a tiny departure
from the $\CC$CDM case since the vacuum energy density itself is evolving
mildly owing to the nonzero value of $\nu$. {For further interesting
thermodynamical considerations about this type of universes (starting from
a de Sitter phase) which show their viability from the point of view of
the generalized second principle of thermodynamics \cite{MimosoPavon2013}.
It is worth noting that several models starting from a de Sitter phase
(deflation) induced by gravitational particle creation of relativistic
particles have also been discussed in the literature \cite{GPC}. As occurs
in the present scenario, some of them also evolve between two extreme de
Sitter phases \cite{LBC12} and the general thermodynamic analysis
presented in \cite{MimosoPavon2013} remains valid (see also \cite{LLC13}
for a possible connection between scenarios driven by decaying  $\Lambda$
models and gravitationally induced particle creation).}

\begin{figure*}\label{fig2}
\centerline{\epsfig{file=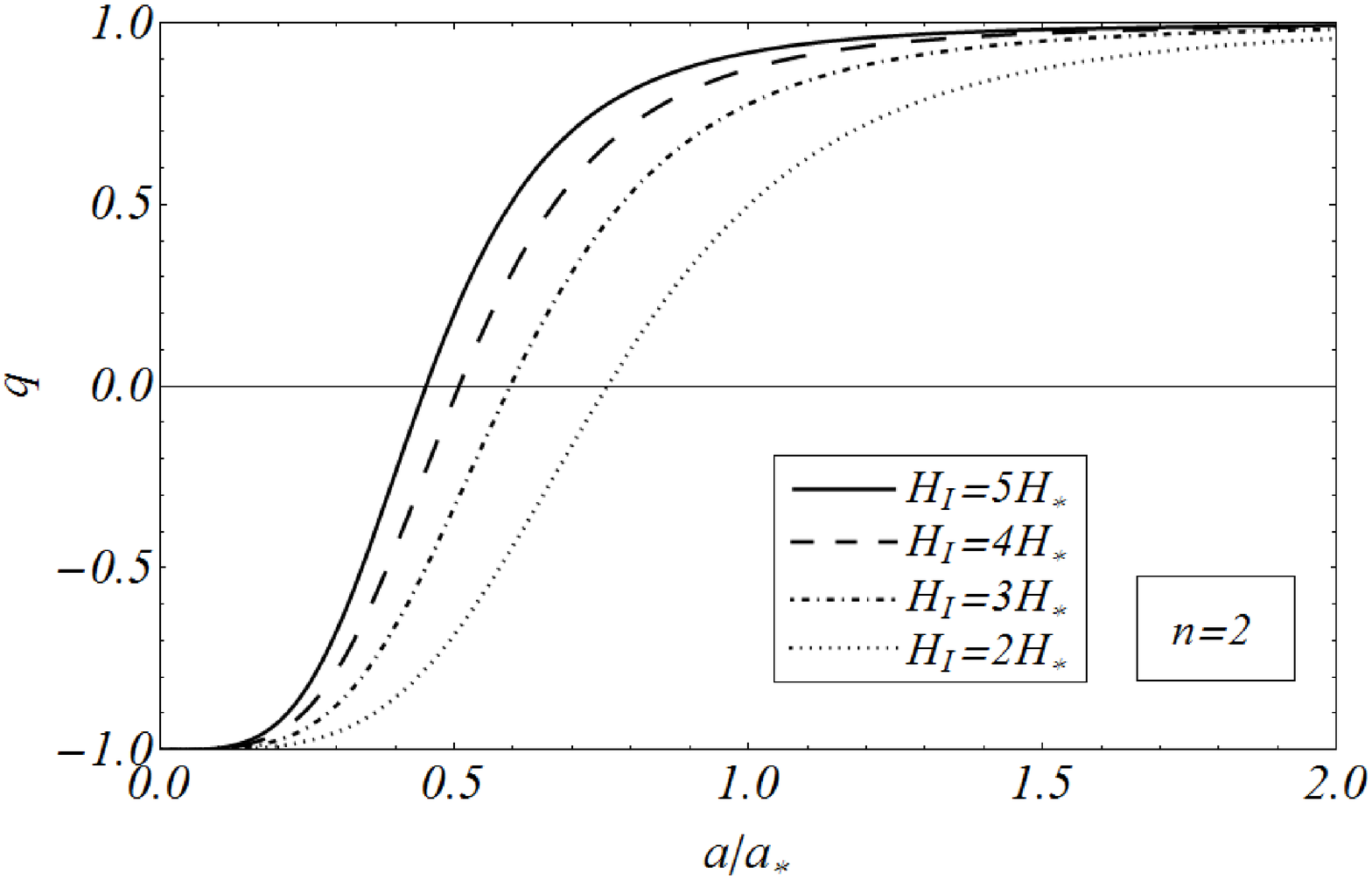,width=6cm}\hspace{0.1cm}\epsfig{file=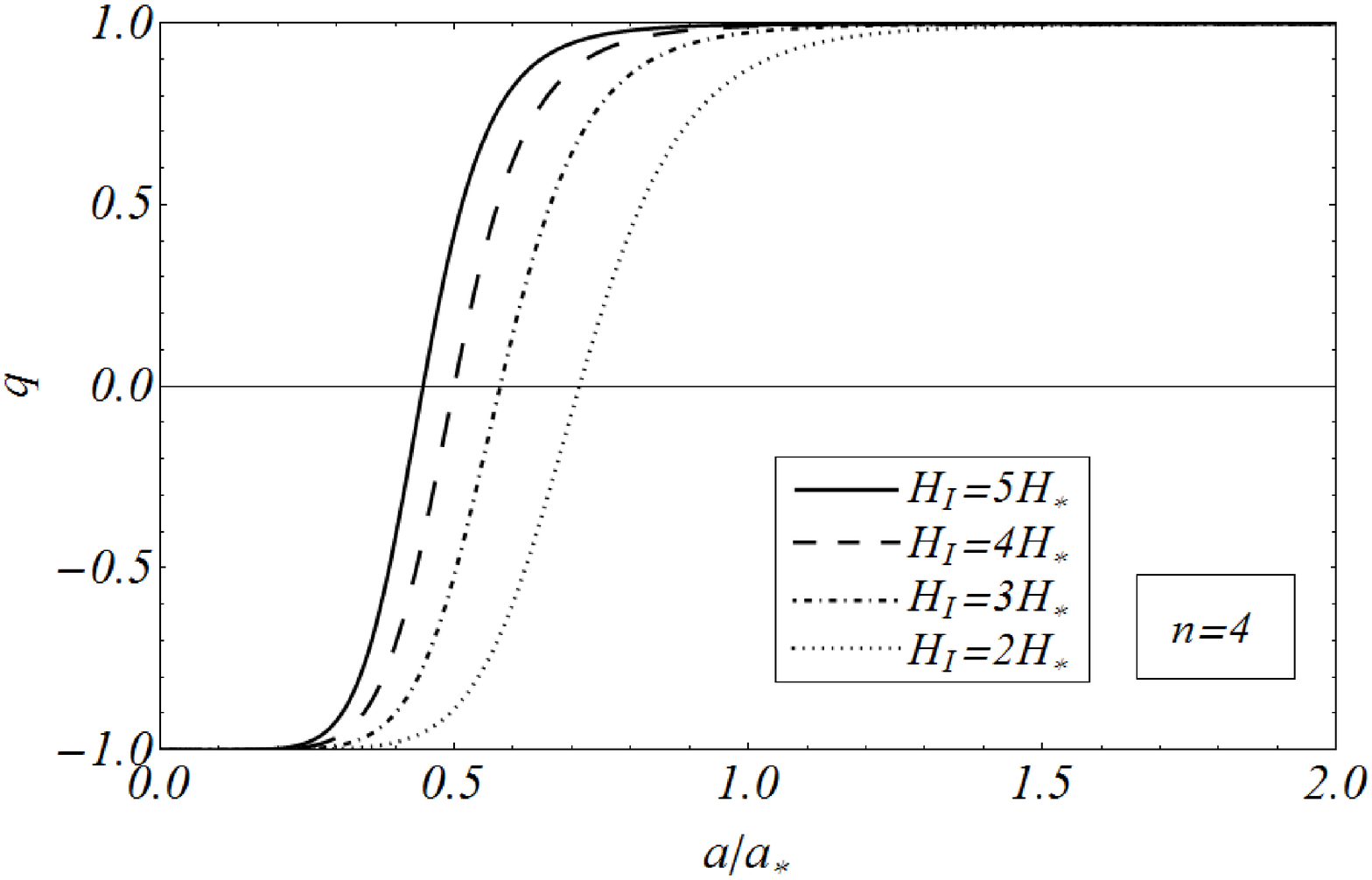,width=6cm}\hspace{0.1cm}\epsfig{file=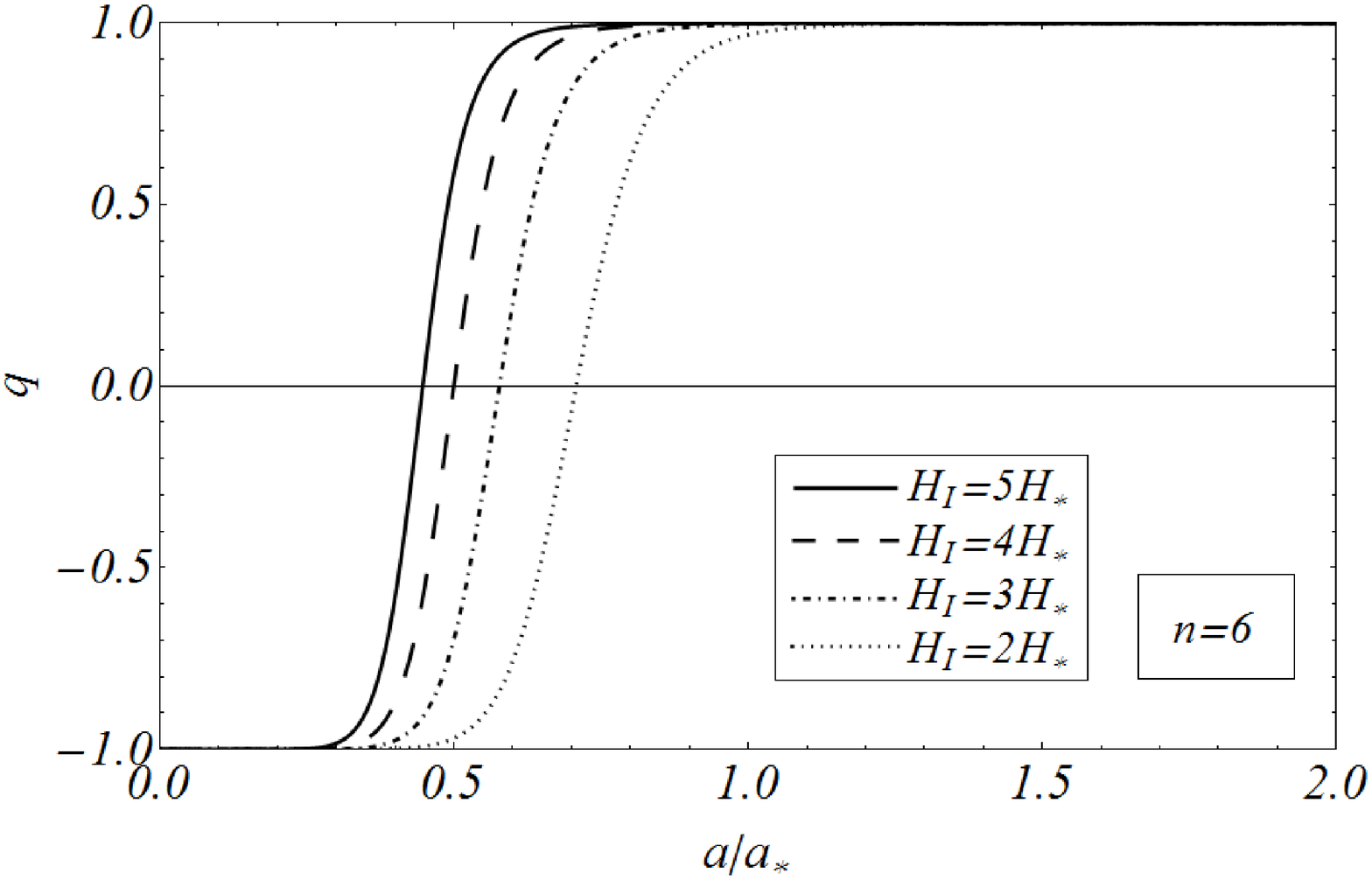,width=6cm}\hspace{0.1cm}}
\caption{Evolution of the decelerating parameter during the primordial era
for several values of the free parameter $n$. All plots were  obtained for
$H^2\gg c_0$ and show the universality of the transition between the early
accelerated de Sitter stage ($q\simeq -1$) and the subsequent decelerated
radiation era ($q\simeq 1$) as driven by the $H^{n+2}$ decaying vacuum
models (from the left to the right we have fixed, respectively, n=2,4,6;
{equivalently, $k=4,6,8$ in Eq.\,(\ref{lambda})}. Note that the transition
occurs faster for the bigger values of the inflationary energy scale $H_I$
and $n$. This general behavior does not change appreciably for any finite
value of $n\geq 1$.}
\end{figure*}

\subsection*{Primordial transition: From an accelerating vacuum to a decelerating radiation phase}
In this case ($c_0\ll H^2$ and for $\omega=1/3$) the deceleration
parameter follows from (\ref{q}) and (\ref{HS1}):

\begin{equation}
q(a)=\frac{(1-2\nu)D\,a^{2n(1-\nu)}-1}{D\,a^{2n(1-\nu)}+1}\,.
\end{equation}
It varies from $ q_I\approx-1$ when $a\to0$ to a positive value near
the standard radiation regime ($q= 1-2\nu\simeq 1$) when
$Da^{2n(1-\nu)} \gg 1$. The primordial transition (pt) between the
early accelerating period and the decelerating radiation phase (when
still $H\gg H_0$) occurs for the scale factor:

\begin{equation}\label{eq:pt}
 a_{pt}=\left[\frac{1}{(1-2\nu)D}\right]^{1/2n(1-\nu)}\,.
\end{equation}
{In Fig. 3} we display some numerical examples of the evolution of
$q(a)$ in this period.


\section{Alternative description in terms of the effective potential}

In Sec. \ref{sect:runningV} we have elaborated on the motivation of the
present model within the general structure of the effective action of QFT
in curved spacetime, and we have used the RG equation (\ref{seriesRLH})
which naturally leads to the expression of the unified model of the vacuum
energy density, Eq.\,(\ref{lambda}). Although at the moment we cannot
provide the effective action leading to this kind of framework in the
general case\,\cite{ShapSol09}, except in some particular
formulations\,\cite{Fossil07}, we can mimic it through an effective scalar
field ($\phi$) model\,\cite{MaiaLima02}. Let us note that any
time-evolving vacuum energy density model can be described in this
way\,\cite{SolaStefancic05}. This can be useful for the usual
phenomenological descriptions of the DE, and can be obtained from the
usual correspondences:
$\rho_{\rm T} \rightarrow \rho_{\phi} =\dot{\phi}^{2}/{2} + V(\phi)$ and
$p_{\rm T} \rightarrow p_{\phi} =\dot{\phi}^{2}/{2} - V(\phi)$
in Friedmann's Eqs.\,(\ref{EE})-(\ref{EE2}). We find ${4\pi
G}\dot{\phi}^{2}=-\dot{H}$ and
\begin{equation}
\label{Vz}
V_{\rm eff}(a)=\frac{3H^{2}}{8\pi G}\left( 1+\frac{\dot{H}}{3H^{2}}\right)=\frac{3H^{2}}{8\pi G} \left( 1+\frac{1}{3}\frac{d\ln H}{d\ln a}\right) \,.
\end{equation}

The effective potential can be readily worked out for our model starting
from  the expression of the Hubble function in the early universe
(\ref{HS1}). We perform the calculation neglecting the small ${\cal
O}(\nu)$ corrections, as they are not important for the present
discussion. The final result is the following:
\begin{equation}\label{eq:Vngeneral}
V_{\rm eff}(a)=\frac{\rho_I}{\alpha^{2/n}}\;\frac{1+Da^{2n}/3}{(1+Da^{2n})^{(n+2)/n}}\,,
\end{equation}
where $\rho_I\equiv 3H_I^2/8\pi G$. The interesting case $n=2$,
corresponding to having a term $H^4$ in the high energy sector of the
vacuum energy density (\ref{lambda}), yields
\begin{equation}
\left.V_{\rm eff}(a)\right|_{n=2}=\frac{\rho_I}{\alpha}\;\frac{1+Da^{4}/3}{(1+Da^{4})^{2}}.
\end{equation}
This specific form was first derived in \cite{LBS12,BLS13}, and is just a
particular case of the general effective potential (\ref{eq:Vngeneral}).
From the general expression it becomes clear that the potential energy
density remains constant, $V_{\rm eff}\sim \rho_I/\alpha$, while $a\ll
D^{-1/(2n)}$ (i.e. before the transition from inflation to the
deflationary regime). However, when the transition is left well behind
(i.e. when $a\gg D^{-1/(2n)}$) the effective potential
(\ref{eq:Vngeneral}) decreases in the precise form $V(a)\sim a^{-4}$,
valid for all $n$, as it should be in order to describe a
radiation-dominated universe independently of the value of $n$. This
result corroborates, in the effective scalar field language, the transit
of the de Sitter stage into the relativistic FLRW regime, which we have
described previously in the original Einstein picture, and shows once more
that our unified model leads to the correct radiation-dominated epoch for
any value of $n$. In other words, the entire class of $\CC(H)$ models
(\ref{lambda}) leads to an acceptable solution of the graceful exit
problem.

\section{From the matter to the residual vacuum domination}\label{sect:MatterEraDE}
In the following we consider the expanding universe well after the
inflationary period and the radiation epoch. To be more precise, we
address the universe at a time after recombination, therefore consisting
of dust ($\omega=0$) plus the running vacuum fluid described by
\eqref{lambda} with $H\ll H_I$. In this case the $H^k$ term ($k\geq 3$) is
completely negligible compared to $H^2$ and that equation reduces to
\begin{equation}\label{GeneralPS}
\Lambda(H)=\Lambda_0+3\,\nu\,(H^2-H_0^2)\,,
\end{equation}
where $\Lambda_0\equiv c_0+3\nu\,H_0^2$ is the current value of the CC.
Obviously, $c_0$ plays an essential role to determine the value of
$\Lambda$, whereas the $H^2$ dependence gives some remnant dynamics even
today, which we can use to fit the parameter $\nu$ to observations. Using
a joint likelihood analysis of the recent supernovae type Ia data, the CMB
shift parameter, and the baryonic acoustic oscillations one finds that the
best fit parameters for a flat universe are: $\Omega_{m0}\simeq 0.27-0.28$
and $|\nu|={\cal O}(10^{-3})$ (see \cite{BPS09,GrandeET11,BPS12}). It is
remarkable that the fitted value of $\nu$ is within the theoretical
expectations when this parameter plays the role of $\beta$-function of the
running CC. As already mentioned, in specific frameworks one typically
finds $\nu=10^{-5}-10^{-3}$ \cite{Fossil07}.

For $H\ll H_I$ and $\omega=0$ the evolution equation of the Hubble
parameter \eqref{HE} becomes simplified. Trading the cosmic time by the
scale factor, upon using $d/dt=aH\,d/da$, it can be rewritten as
\begin{equation}
 a\,H\,H^\prime+\frac{3}{2}(1-\nu)H^2-\frac{c_0}{2}=0\,,
\end{equation}
where the prime denotes derivative with respect to the scale factor $a$.
The above equation can now be integrated with the result (\ref{eq:Ha})
(see the Appendix)

\begin{equation}
\label{eq:Ha1}
{H}^{2}(a) = \frac{H_0^2}{1-\nu} \left[(1-\Omega_{\CC}^{0})\,a^{-3(1-\nu)}+\Omega_{\Lambda}^0-\nu \right]\,,
\end{equation}
where we have used  the corresponding boundary condition at the present
time: $c_0=3H_0^2(\Omega_{\Lambda}^0-\nu)$. Notice that the previous
equation can, if desired, easily be reexpressed in terms of the redshift
$z$ through the relation  $1+z=1/a$.

Similarly, the matter and vacuum energy densities are found to be (see
Appendix):
\begin{equation}\label{mRGa1}
\rho_m(a) =\rho_m^0\,a^{-3(1-\nu)}\,,
\end{equation}
and
\begin{equation}\label{CRGa1}
\rL(a)=\rLo+\frac{\nu\,\rho_m^0}{1-\nu}\,\left[a^{-3(1-\nu)}-1\right]\,.
\end{equation}
where $\rho_m^0$ and $\rLo$ are the corresponding values at present
($a=1$). The total energy density reads
\begin{equation}\label{eq:rhototalNOW}
\rho_{\rm T}(a)=\frac{\rho_m^0}{1-\nu}\left[a^{-3(1-\nu)}-\nu\right]+\rLo\,.
\end{equation}

Integrating once more the equation \eqref{eq:Ha1} with respect to the
cosmic time we obtain the following time dependence of the scale factor:
\begin{equation}\begin{split}
a(t)=&\left(\frac{1-\Omega_{\Lambda}^0}{\Omega_{\Lambda}^0-\nu}\right)^{\frac{1}{3(1-\nu)}} \times\\
     &\sinh^{\frac{2}{3(1-\nu)}} \left[3H_0\sqrt{(1-\nu)(\Omega_{\Lambda0}-\nu)}t/2\right]\,.
\end{split}\end{equation}
As expected, for $\nu\ll1$ at late enough times the above solution mimics
the Hubble function $H(a)$ of the usual flat $\Lambda$-cosmology, which
means that the final dynamics of the universe is determined by a single
parameter namely $\Omega_{\Lambda}^0$ or $\Omega_m^0$, which are well
known to be related by the cosmic sum rule
$\Omega_m^0+\Omega_{\Lambda}^0=1$.

From these equations it is clear that for $\nu=0$ we recover exactly the
$\Lambda$CDM expansion regime, the standard scaling law for
nonrelativistic matter and a strictly constant vacuum energy density
$\rho_\Lambda=\rho_{\Lambda}^0$ (hence $\Lambda=\Lambda_0$). Recalling
that $|\nu|$ is found to be rather small when the model is confronted with
the cosmological data, $|\nu|\leq{\cal O}(10^{-3})$
\cite{BPS09,GrandeET11}, we see that the model under consideration
deviates very small from the $\CC$CDM, specially in the postinflationary
epoch, where the only distinctive trace left of the model is the existence
of a slowly evolving vacuum energy density or cosmological term
(\ref{GeneralPS}). This is compatible with the general notion of dynamical
dark energy, which in this case would be caused by a dynamical vacuum in
interaction with matter.

At very late time we get an effective cosmological constant dominated era,
$H\approx H_0\,\sqrt{(\Omega_{\Lambda}^0-\nu)/(1-\nu)}$, see
Eq.\,\eqref{eq:Ha1} for sufficiently large $a$, that implies a pure de
Sitter phase of the scale factor. This is the late time de Sitter phase or
DE epoch.

\subsection*{The deceleration parameter in recent times}
In the epoch under consideration, we have $\omega=0$ and $H/H_I\ll 1$.
Thus, with the help of Eqs. \eqref{q} and (\ref{eq:Ha1}) the deceleration
parameter takes the form
\begin{equation}
q(a)=\frac{(1-3\nu)\Omega_m^0 a^{-3(1-\nu)}-2(1-\nu-\Omega_m^0)}{2\Omega_m^0 a^{-3(1-\nu)}+2(1-\nu-\Omega_m^0)}\,,
\end{equation}
In the limit $\nu\to 0$ this expression reduces to that of the $\CC$CDM
model. In particular, the current value ($a=1$) is
$q_0=(3\Omega_m-2)/2\simeq-0.58$, where $\Omega_m^0\simeq 0.28$. The
late-time transition (lt) -- in contrast to the aforementioned primordial
transition (\ref{eq:pt}) -- between the decelerated matter-dominated era
and the late accelerated residual vacuum stage of the expanding universe
occurs when
\begin{equation}
 a_{lt}=\left[\frac{(1-3\nu)\Omega_m}{2(1-\nu-\Omega_m)}\right]^{1/3(1-\nu)}\,.
\end{equation}
In the limit $\nu\to 0$ it gives $a_{lt}\simeq0.58$. {In Fig.\,4 we show}
this late transition point and compare it with the slightly different
values obtained for the case when $\nu\neq 0$.

\begin{figure}\label{fig4}
\epsfig{file=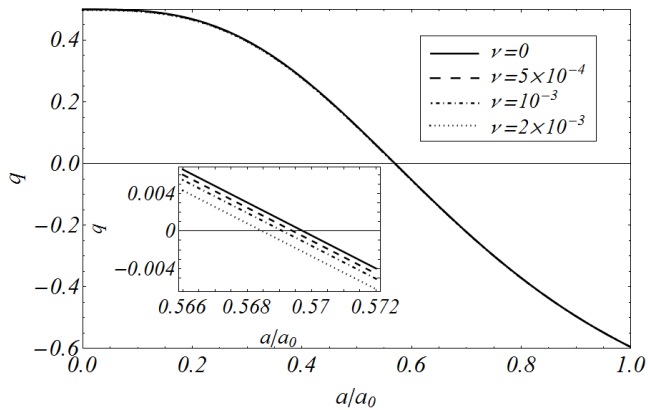,width=8.5cm}
\caption{Evolution of the decelerating parameter during the
late stages, when $H\ll H_I$. This figure shows the small
departure of the current model (with $\nu\lesssim 10^{-3}$) from
the $\Lambda CDM$ model. The effect of greater values of $\nu$ is
summarized in shifting forward in time the transition point from
deceleration to acceleration into the current accelerated stage.}
\end{figure}

Despite the dynamical character of the vacuum energy (\ref{GeneralPS})
near our time, it is important to understand that a model of this kind
would not work for $c_0=0$, i.e. with only pure $H$-dependent terms on
$\Lambda$. This has been proven in \cite{BPS09,BPS12} and recently
discussed also in \cite{XuET11}. The basic drawback of the $c_0=0$ models
is that the deceleration parameter never changes sign, and therefore the
universe always accelerates or always decelerates\,\cite{BPS12}. In the
present case this can be seen as follows. We can easily check that the
condition $c_0=0$ enforces Eq.\,(\ref{eq:Ha1}) to take the simpler form
$H^2(a)=H_0^2\,a^{-3(1-\nu)}$. From here we immediately find
\begin{equation}\label{eq:qnu}
q=-1-a\,\frac{H'(a)}{H(a)}=-1+\frac32\,\left(1-\nu\right)\,.
\end{equation}
It follows that acceleration ($q<0$) is possible only for $\nu>1/3$, which
is unacceptable since we have emphasized that $|\nu|\ll 1$. What is more,
since $q$ given by (\ref{eq:qnu}) is a constant (i.e. independent of time
or of the scale factor) it can only have a sign for a given value of
$\nu$, so even if we would admit $\nu>1/3$ as a mere phenomenological
possibility, we would be also admitting that the universe has been
accelerating forever, which is of course difficult to accept.


\subsection*{The present value of the vacuum energy}

After showing the importance of having a nonvanishing $c_0$ term in
our unified vacuum model $\CC(H)$, Eq.\, (\ref{lambda}), specially
for the low energy segment of the cosmological observations, let us
note that the RG formulation of it (cf. Sect. \ref{sect:runningV})
provides a natural explanation for the presence of such $c_0\neq 0$
value, to wit: the integration of the RG equation (\ref{seriesRLH})
must necessarily lead to a nonvanishing additive term in the
structure of $\rL(H)$. Therefore, a term of this sort is naturally
motivated in this framework. From it the current value of the vacuum
energy density reads $\rLo=(c_0+3\,\nu\,H_0^2)/(8\pi\,G)$. Of course
the value of $c_0$ must be fixed by the boundary condition of the RG
differential equation, which is fixed by current observations:
$\rLo=\rL(H_0)$.

The following observation is now in order: despite our model providing a
dynamical explanation for the drastic reduction of the early vacuum energy
of our universe from $\rL(H_I)$ to the comparatively very small quantity
$\rLo\ll\rL(H_I)$, and at the same time insuring that $\rL(H)$ will be
totally harmless for the correct onset of the radiation epoch (see
Sect.\,\ref{sect:deSitterRadiation}),  the ultimate value that $\rL(H)$
takes at present, i.e. $\rLo$, cannot be predicted within the model itself
and hence can only be extracted from observations. Notice that if we could
have the ability to predict this value it would be tantamount to solve the
CC problem\,\cite{Weinberg89}. This is of course the toughest part of the
longstanding  unsolved cosmological constant problem. In our case,
however, we have ascribed a new look to the problem, one that could
perhaps make it more amenable for an eventual solution; namely, we have
shown that the cosmological term which we have measured at present is not
the same immutable tiny quantity that the $\CC$CDM assumes for the entire
cosmic history, but rather a time-evolving variable that underwent a
dramatic dynamical reduction from the inflationary time until the present
days.


\section{Conclusions}
In this article we have proposed a new phenomenological scenario which
provides a complete cosmic expanding history of the universe. It is based
on a dynamical model (in fact an entire class of models) for the vacuum
energy that covers all the relevant states of the cosmic evolution. The
function $\CC=\CC(H)$ that we propose involves a power series of the
Hubble rate $H$, which in practice consists of an additive term, a power
$H^2$ and finally a higher power $H^k$ ($k>2$) which is responsible for
the transition from the inflationary stage to the FLRW radiation epoch.
The ansatz that we used is motivated by the covariance of the effective
action of QFT in curved spacetime and in this sense the even powers of $H$
are preferred, although for completeness we have described the general
case.

First of all the model itself predicts that the universe starts from a
nonsingular state and thus we can solve easily the horizon problem. {This
early accelerated regime associated with the inflation has a natural
ending by virtue of the faster decrease of the vacuum energy density
thereby generating the radiation fluid and the ultrarelativistic gas
particles.} The novelty in the current work is the fact that the dynamical
vacuum model which we propose smoothly accommodates the standard cosmic
epochs characteristic of the $\Lambda$CDM model, namely the
radiation-dominated, matter-dominated and late-time de Sitter phase
($\Lambda=const.$). The universe described in our proposal therefore
evolves from a primeval de Sitter epoch to another late time de Sitter
epoch, which is the one we have recently entered. Let us note that the
mechanism for inflation in our case is quite different from that of usual
inflaton models. In this sense it may provide an alternative to them,
especially after realizing that the PLANCK results\,\cite{Ade13} {rule out
some of these scalar field models}, whereas in our case the sustained
plateau we have in the vacuum inflationary phase could perhaps help
explain better the new data and in particular the so-called ``unlikeliness
problem''\,\cite{Steinhardt2013}. A devoted analysis is of course needed,
but it is clear that we remain as motivated as ever to look for new ideas
and alternative mechanisms for inflation. Let us finally note that our
model, apart from avoiding the initial singularity and alleviating the
horizon and graceful exit problems, it also helps to mitigate the
cosmological constant problem i.e., the fact that the observed value of
the vacuum energy density ($\rho_{\Lambda}=c^{2} \Lambda /8\pi G\simeq
10^{-47}\,GeV^4$) is many orders of magnitude below the value found using
quantum field theory.



\acknowledgements

ELDP is supported by a fellowship from CNPq and JASL is partially
supported by CNPq and FAPESP (Brazilian Research Agencies). SB acknowledges support by the Research Center for Astronomy of the Academy of Athens
in the context of the program {\it ``Tracing the Cosmic Acceleration''}. JS has been supported in part by projects
FPA2010-20807 and CPAN (Consolider CSD2007-00042) and also by 2009SGR502
Generalitat de Catalunya.

\appendix

\section{General Solutions}
\subsection{From the early de Sitter stage to the $\omega$-dominated phase}

At early stages of the universe, the $c_0$ parameter is negligible and the
Eq.\,\eqref{HE} for the evolution of the Hubble function becomes
\begin{equation}
\dot H+\frac{3}{2}(1+\omega)H^2\left[1-\nu-\alpha\left(\frac{H}{H_I}\right)^n\right]=0\,.
\end{equation}
The integration of the above equation gives
\begin{equation}\label{HS1a}
 H(a)=\frac{\tilde H_I}{\left[1+D\,a^{n\xi}\right]^{1/n}}\,,
\end{equation}
where $\xi\equiv3(1+\omega)(1-\nu)/2$ and $\tilde H_I\equiv
H_I[(1-\nu)/\alpha]^{1/n}$. We  stress that in our analysis we
consider epochs of the cosmic evolution where matter is dominated by
the relativistic or the nonrelativistic components, i.e. epochs
where we have $\omega=1/3$ and $\omega=0$ respectively, without
considering the interpolation regime between the two. Therefore, in
practice for all the considerations in this section, we have
$\omega=1/3$ -- and so $\xi=2(1-\nu)$ -- as our discussion is
related to the transition from the initial de Sitter to the
radiation dominated universe. However, a simulation of the
$\omega$-dependence from $\omega=0$ to $\omega=1/3$ is done in {Fig.
2}.

In  Eq.\,(\ref{HS1a}), $D$ is an integration constant that can be fixed
using the condition $H(a_\star)\equiv H_\star$ (where
$a_\star=a(t_\star)$, typically corresponding to the initial time
$t_\star$ of the $\omega$-fluid dominated era). Thus,
\begin{equation}D=a_\star^{-n\xi}\left[\left(\frac{\tilde H_I}{H_\star}\right)^n-1\right]\,,\end{equation}
and it is greater than zero for $\tilde H_I>H_\star$. Note
that if $D=0$ the solution remains always de Sitter.

Using the auxiliary variable
\begin{equation}
u=-\frac{1}{D\,a^{n\xi}}\,,
\end{equation}
which transforms Eq.\,\eqref{HS1a} as
\begin{equation}
\dot u=-n\xi \tilde H_I u^{1+1/n}\left(u-1\right)^{-1/n}\,,
\end{equation}
and its inversion results:
\begin{equation}
\frac{d t}{d u}=-\frac{1}{n\xi \tilde H_I}u^{-(1+1/n)}\left(u-1\right)^{1/n}\,.
\end{equation}
The second derivative may be put in the form:
\begin{equation}
u(1-u)\frac{d^2t}{d u^2}+\left[1+\frac{1}{n}-u\right]\frac{d t}{d u}=0\,.
\end{equation}
Hence, we have the hypergeometric equation with parameters $a=0$, $b=1/n$,
and $c=1+1/n$. Its integration yields
\begin{equation}
t(u)=B-A\,n\,u^{-1/n}F\left[-\frac{1}{n},-\frac{1}{n},1-\frac{1}{n},u\right]\,,
\end{equation}
where $B$ and $A$ are integration constants. We can set $B=0$ if the
origin of time is placed just after the inflation period and $t$ is then
the cosmic time in the FLRW epoch.  Using Euler's relation for the
hypergeometric function and the boundary condition (when $t=t_\star$ at
the end of the inflationary period) for the Hubble parameter $H$ the above
solutions can be rewritten as:
\begin{equation}\label{tna}
t(a)=B+\frac{\left(1+D\,a^{n\xi}\right)^{\frac{1+n}{n}}}{\xi\,\tilde H_I\,D\, a^{n\xi}} F\left[1\,,1\,,1-\frac{1}{n}\,,\frac{-1}{D\,a^{n\xi}}\right]\,,
\end{equation}
and for $n=2$ this solution becomes
\begin{equation}\begin{split}
 t(a)=&B+\frac{1}{\xi H_I}\sqrt{\frac{\alpha\left(1+D\,a^{2\xi}\right)}{1-\nu}}\\
      &-\frac{1}{\xi H_I}\sqrt{\frac{\alpha}{1-\nu}}\text{ArcCoth}\sqrt{1+D\,a^{2\xi}}\,.
\end{split}\end{equation}

Using the Einstein equations and the above solutions we can obtain the corresponding energy densities:
\begin{equation}\label{rho_1la}
  \rho_\Lambda(a)=\tilde\rho_I\frac{1+\nu\,D\, a^{n\xi}}{\left[1+D\,a^{n\xi}\right]^{1+2/n}}\,,
\end{equation}
\begin{equation}\label{rho_1a}
 \rho(a)=\tilde\rho_I\frac{(1-\nu)D\,a^{n\xi}}{\left[1+D\,a^{n\xi}\right]^{1+2/n}}\,,
\end{equation}
\begin{equation}\label{rho_1ta}
 \rho_{\rm T}(a)=\tilde\rho_I\frac{1}{\left[1+D\,a^{n\xi}\right]^{2/n}}\,,
\end{equation}
with $\tilde\rho_I\equiv3\tilde H_I^2/8\pi G$. It is easy to check that
these expressions correctly reproduce the energy densities we have used in
Sec. \ref{sect:deSitterRadiation} for the primeval de Sitter and radiation
dominated epochs.

\subsection{From the $\omega$-dominated era to the residual vacuum stage}

Next we consider the derivation of the corresponding formulas for the more
recent universe when the $\omega$-fluid plus a vacuum fluid [described by
\eqref{lambda}] expand under he condition $H\ll H_I$. In this case the
evolution equation for the Hubble parameter Eq.\,\eqref{HE} can be
approximated as

\begin{equation}\label{radiationa}
 a\,H\,H^\prime+\xi H^2-\frac{(1+\omega)}{2}c_0=0\,,
\end{equation}
where the prime denotes derivative with respect to the scale factor $a$,
and again $\xi\equiv3(1+\omega)(1-\nu)/2$. The first integral of this
equation gives
\begin{equation}\label{Ha}
 H^2=\frac{c_0}{3(1-\nu)}\left[\left(\frac{C_1}{a}\right)^{2\xi}+1\right]\,,
\end{equation}
where the constant
\begin{equation}
{C_1}^{2\xi}={a_0}^{2\xi}\left[\frac{3H_0^2(1-\nu)}{c_0}-1\right]\,,
\end{equation}
is obtained from the condition $H(a_0)\equiv H_0$ today.

Using the above solutions, the Friedmann equations provide the total and
the $\omega$-fluid densities
\begin{equation}\label{rho_2t}
 8\pi G\rho_T(a)=\frac{c_0}{1-\nu}\left[\left(\frac{C_1}{a}\right)^{2\xi}+1\right]\,,
\end{equation}
\begin{equation}\label{rho_2}
 8\pi G\rho(a)=c_0\left(\frac{C_1}{a}\right)^{2\xi}\,.
\end{equation}
In a more explicit form, the Hubble function (\ref{Ha}) reads
\begin{equation}
\label{eq:Ha}
 {H}^{2}(a) = \frac{H_0^2}{1-\nu} \left[\Omega_{X}^{0}\,a^{-2\xi}+\Omega_{\Lambda}^0-\nu \right]\,,
\end{equation}
where we have the sum rule $\Omega_{X}^{0}+\Omega_{\Lambda}^0=1$, and we
have set $\omega=0$ ($X=m$) since we are in the matter-dominated epoch.
The $\omega$-fluid density (\ref{rho_2}) can be expressed as
\begin{equation}\label{mRG}
\rho(a) =\rho^0\,a^{-2\xi}\,,
\end{equation}
where $\rho^0$ is the current value. We can see that for $\nu=0$ we
retrieve the standard scaling $\rho=\rho^0\,a^{-3(1+\omega)}$. The departure
from this law caused by a nonvanishing $\nu$ is related to the exchange of
energy between matter and vacuum. By the same token the vacuum is no
longer static, and the effective CC evolves as
\begin{equation}\label{rho_2l}
 \Lambda(a)=\frac{c_0}{1-\nu}
\left[\nu\left(\frac{C_1}{a}\right)^{2\xi}+1\right]\,.
\end{equation}
The corresponding vacuum energy density is the following:
\begin{equation}\label{CRG}
\rL(a)=\rLo+\frac{\nu\,\rho^0}{1-\nu}\,\left[a^{-2\xi}-1\right]\,.
\end{equation}
We see that only for $\nu=0$ we recover $\CC=c_0=$const. and
$\rL(a)=\rLo=$const., as in the $\CC$CDM case. Furthermore,  we can easily
check that Eqs.\,(\ref{mRG}) and (\ref{CRG}) satisfy the overall local
conservation law (\ref{lambdavar}), which can be rewritten in terms of the
scale factor as follows:
\begin{equation}\label{Bronstein2}
\rho'_\CC(a)+\rho'(a)+\frac{3}{a}(1+\omega)\,\rho(a)=0\,,
\end{equation}
where the prime indicates differentiation with respect to the scale
factor.

We can integrate Eq. \eqref{Ha} to obtain the time evolution of the scale
factor $a(t)$:
\begin{equation}
\label{a_fa}
 a(t)=C_1\sinh^{1/\xi}\left[\sqrt{3c_0(1-\nu)}(1+\omega)(t-C_2)/2\right]\,.
\end{equation}
Without losing generality we can set $C_{2}=0$. Substituting (\ref{a_fa})
in the previous equations we immediately get the time-evolving functions
$\rho=\rho(t)$ and $\CC=\CC(t)$.

Let us finally mention for completeness that there are cases where we have
to deal with a mixture of cold matter and radiation.
Defining $\Omega_{m}^{0}$ and $\Omega_{r}^{0}$ as the standard
nonrelativistic and radiation density parameters at the present time, one
can show that the complete Hubble function reads
\begin{equation}
H^{2}(a)= \frac{H_0^2}{1-\nu} \left[
\Omega_{m}^{0}a^{-3(1-\nu)}+\Omega_{\Lambda}^0+
\Omega_{r}^{0} a^{-4(1-\nu)}-\nu \right]\,,
\end{equation}
where the density parameters satisfy the extended sum rule
$\Omega_{m}^{0}+\Omega_{r}^{0}+\Omega_{\Lambda}^0=1$.


\end{document}